\documentstyle[revtex,eqsecnum]{aps}
\begin{document}

\draft

\begin{title}
Pressure and density of vacancies in solid $^4$He
\end{title}
\author{ Dirk~Jan~Bukman }
\begin{instit}
 Department of Physics,
 Simon Fraser University,\\
 Burnaby,
 British Columbia,
 Canada V5A 1S6
\end{instit}
\author{ J.M.J.~van~Leeuwen}
\begin{instit}
 Instituut-Lorentz,
 Rijksuniversiteit te Leiden,\\
 P.O.~Box 9506,
 2300 RA Leiden,
 The Netherlands
\end{instit}
\begin{abstract}
\noindent Crystals of $^4$He contain vacancies that move around by a
quantum mechanical hopping process. The density and pressure of these
vacancies can be experimentally studied.
The accuracy of the experiments
is high enough to detect the effect of the Bose statistics of the
vacancies. In this paper we examine the effect of the
hard-core repulsion
between the vacancies, which should also have a measurable effect
on their behaviour. We set up a virial expansion for a lattice gas of
hard-core particles, and calculate the second virial coefficient.
It turns out that the vacancies behave as ideal Bose particles at
low temperatures, but that the hard-core interaction makes them behave
more and more like fermions as the temperature increases.
\end{abstract}
\narrowtext

\section{Introduction}

The thermodynamic behaviour of vacancies in solid $^4$He is an
interesting experimental and theoretical problem \cite{And1}.
Vacancies
in Helium are more mobile than in any other solid. At the temperatures
where solid Helium exists the motion of vacancies requires a quantum
mechanical description. They hop from site to site with a certain
rate $\nu_v$,
leading to a band of
states $\varepsilon(\vec{k})$, much like the electron motion in
the tight binding approximation. For $^4$He the vacancies are
obviously
bosons, since the creation operator for a vacancy is the annihilation
operator for a $^4$He particle, which is a boson \cite{And2}.
The fact that two vacancies cannot
occupy the same lattice site has to be incorporated as a hard-core
potential for the hopping bosons.
So the simplest model for vacancies in $^4$He is that of a gas of
hard-core bosons on a lattice. In reality the strain fields
around the
vacancies produce a more complicated interaction between them than
the simple on-site exclusion. However, we consider the hard-core
boson approximation as a sufficiently realistic description
of the vacancy motion to leave out these further refinements,
in order not to complicate the model too much.
The hard-core bosons on a lattice are known to be equivalent to
a spin-$\frac{1}{2}$ quantum system with an interaction of the
XY type
\cite{M+M}. It is however not this analogy that is exploited in
this paper.
The reason is that the vacancy system is in practical circumstances
always extremely dilute, which corresponds in the spin analogy to
a system in an extremely large external magnetic field. In this limit
the spin analogy is of little use and in our opinion less transparent
than the particle language.

Experiments show that the percentage of vacancies in a crystal
is at most of the order of 1\% \cite{Goodk}, at least at temperatures
of the order of 1K, where the experiments take place.
As a first approximation the vacancies thus behave as an
ideal gas. However, present day experiments are sufficiently
accurate that
effects of Bose statistics can be detected. It is one of our points
that then also effects of the hard-core interaction become detectable.

In this paper we present a systematic analysis of the vacancies
using $\exp (-\Delta/k_BT)$ as a small parameter, $\Delta$ being the
excitation energy, or band gap, required to create a vacancy.
This is equivalent
to a virial expansion \cite{Huang} for the quantum lattice gas,
and we work out
the properties in detail up to the second virial coefficient.
A general formula for the second virial coefficient of the hard-core
Bose lattice gas is derived and evaluated for hypercubic lattices
in one, two, and three dimensions. The case of the hcp lattice, which
applies to real solid $^4$He, will be treated elsewhere.

\section{The hard-core Bose lattice gas}

The vacancies are represented by Bose creation and annihilation
operators $b_i^{\dagger}, b_i$ obeying the usual Bose commutation
relations. The Hamiltonian for the vacancies is given by
\FL
\begin{equation}
{\cal H} = -t \sum_{\langle i,j \rangle} ( b_i^{\dagger}b_j +
b_j^{\dagger}b_i) +(\Delta + ct ) \sum_i b_i^{\dagger} b_i
+\frac{U}{2} \sum_i b_i^{\dagger} b_i^{\dagger} b_i b_i .
\label{ham}
\end{equation}
Here the transfer integral $t$ is equal to $h\nu_v$, and the
hops take place between all pairs of nearest neighbours $\langle
i,j \rangle$ on the lattice. The coordination number of the
lattice is $c$, and $\Delta$ is the energy required to create a
vacancy. It functions as minus the chemical potential for the
vacancies. The last term represents the vacancy-vacancy repulsion.
It
could be omitted in favour of a change to on-site Fermi commutation
relations for the $b_i^{(\dagger)}$, but we prefer to work with
standard commutation relations and a potential $U$ which
penalizes the simultaneous occurrence of two or more vacancies
at the same site. We will let $U\rightarrow \infty$, or $U$ is
much larger than any other energy in the problem.

Without the potential term we have an ideal Bose lattice gas, the
Hamiltonian of which can be diagonalized by the canonical
transformation
\begin{eqnarray}
b(\vec{k}) &=& \frac{1}{\sqrt{N}} \sum_j b_j e^{i\vec{k}\cdot
\vec{r}_j},\nonumber \\
&& \\
b_j &=& \frac{1}{\sqrt{N}} \sum_{\vec{k}} b(\vec{k})
e^{-i \vec{k} \cdot \vec{r}_j},\nonumber
\label{trafo}
\end{eqnarray}
where $N$ is the number of sites in the system. We use periodic
boundary conditions, and the wave vectors $\vec{k}$ are restricted
to the first Brillouin zone of the lattice.

Inserting (\ref{trafo}) into the first two terms of (\ref{ham})
we obtain the unperturbed Hamiltonian
\begin{equation}
{\cal H}_0 = \sum_k [\Delta + \varepsilon(\vec{k}) ] b^{\dagger}
(\vec{k}) b(\vec{k}),
\label{h0}
\end{equation}
where $\varepsilon(\vec{k}) $ is given by
\begin{equation}
\varepsilon(\vec{k}) = t \sum_{\delta} (1-\cos \vec{k}\cdot\vec{r}_
{\delta}),
\label{eps}
\end{equation}
and $\vec{r}_{\delta}$ is the set of the $c$ nearest neighbour
positions with respect to a centrally chosen site. $\Delta$ is
the gap of the energy band, since we have $\varepsilon(\vec{0})
=0$ as lowest energy in the center of the Brillouin zone.
One sees that $t$ is
a measure for the bandwidth, which for a $d$-dimensional
hypercubic lattice is $w=4dt$.

The interaction is written in terms of $b^{(\dagger)}(\vec{k})$
as
\begin{equation}
V=\frac{U}{2N} \sum_{\vec{k}_i,\vec{G}} \delta_{\vec{k}_1+\vec{k}_2,
\vec{k}_3+\vec{k}_4+\vec{G}} b^{\dagger}(\vec{k}_1)b^{\dagger}
(\vec{k}_2)
b(\vec{k}_3) b(\vec{k}_4),
\label{vk}
\end{equation}
where $\vec{G}$ is a vector of the reciprocal lattice.
The matrix elements of the interaction have no other
structure than the conservation of the total incoming and
outgoing momentum (up to a reciprocal lattice vector), a feature
which is of great advantage in solving the two-particle problem.

\section{The virial expansion}

The grand partition function of the vacancy system is given by
\begin{equation}
\Xi = \mathop{\rm Tr}\nolimits e^{-\beta {\cal H}},
\label{xi}
\end{equation}
with ${\cal H}$ given by (\ref{ham}), $\beta=1/k_BT$, and
$\mathop{\rm Tr}\nolimits$ stands
for the trace over all symmetrized states. As ${\cal H}$ conserves the
number of vacancies, $\Xi$ can be expanded as
\begin{equation}
\Xi = \sum_{n=0}^{\infty} Z_n e^{-n\beta \Delta},
\label{expa}
\end{equation}
where $Z_n$ is the canonical partition sum for $n$ vacancies excluding
the contribution from the gap $\Delta$. Of course, $Z_0=1$, and
\begin{equation}
Z_1=\mathop{\rm Tr}\nolimits_1 e^{-\beta({\cal H}_0-\Delta)} =
\sum_{\vec{k}} e^{-\beta
\varepsilon (\vec{k})},
\label{z1}
\end{equation}
because for one vacancy no hard-core effects enter.
$\mathop{\rm Tr}\nolimits_n$ is the trace
over $n$-vacancy states.

For $\ln \Xi$ we may deduce from (\ref{expa})
\begin{equation}
\beta p N v_0 =\ln \Xi =N \sum_{\ell=1}^{\infty} b_{\ell} e^{-\ell
\beta
\Delta},
\label{vircs}
\end{equation}
where $v_0$ is the volume of the unit cell, $p$ is the pressure
of the
vacancies, and the $b_{\ell}$ are the fugacity expansion
coefficients.
The first of these, $b_1$, reads
\begin{equation}
b_1(\beta) = \frac{1}{N} Z_1 = \frac{v_0}{(2 \pi)^d} \int_{BZ}
d\vec{k} \,e^{-\beta\varepsilon(\vec{k})},
\label{b1}
\end{equation}
where we have replaced the sum over $\vec{k}$ by an integral over the
Brillouin zone.
In contrast to the continuum ideal gas, $b_1$ is not simply related
to a thermal wavelength $\lambda$ as
\begin{equation}
b_1=v_0/\lambda^{d}.
\label{b1l}
\end{equation}
A formula of this type only results when the temperature is so low
that
$\varepsilon(\vec{k})$ may be replaced by its low-momentum behaviour
\begin{equation}
\varepsilon(\vec{k}) \approx \frac{t}{2} \sum_{\delta} (\vec{k}
\cdot \vec{r}_\delta )^2 \equiv \frac{\hbar^2 k^2}{2m^*},
\label{lowk}
\end{equation}
defining an effective mass $m^*$ for the vacancies. For
$d$-dimensional
hypercubic lattices we find
\begin{equation}
m^*=\frac{\hbar^2}{2 t a^2},
\label{mstar}
\end{equation}
where $a$ is the lattice constant. Using (\ref{lowk}) in (\ref{b1})
and extending the $\vec{k}$-integral beyond the Brillouin zone to
infinity (as
is allowed for large $\beta$ or small $T$) one finds (\ref{b1l})
with
\begin{equation}
\lambda^2=h^2/2\pi m^* k_BT.
\label{lambda}
\end{equation}

The second term $b_2$ in (\ref{vircs}) is our main concern in this
paper.
We find from (\ref{expa}) and (\ref{vircs})
\begin{eqnarray}
b_2&=&\frac{1}{N}\left( Z_2 - \frac{Z_1^2}{2}\right)\nonumber\\
&=&
\frac{1}{N}\left\{ \mathop{\rm Tr}\nolimits_2 e^{-\beta({\cal H}
-2 \Delta )} -
\frac{1}{2}\left( \mathop{\rm Tr}\nolimits_1 e^{-\beta({\cal H}_0 -
\Delta)}\right)^2\right\}.\nonumber\\
\label{b2}
\end{eqnarray}
We rewrite this expression by adding and subtracting the
contribution of
the non-interacting two-vacancy system. So we define
\begin{equation}
b_2^{\rm int}=\frac{1}{N} \mathop{\rm Tr}\nolimits_2\left[
e^{-\beta({\cal H}-2\Delta)}-
e^{-\beta({\cal H}_0 -2\Delta)}\right]
\label{b2int}
\end{equation}
as the contribution of the hard-core interaction, and
\begin{equation}
b_2^0=\frac{1}{N} \left[\mathop{\rm Tr}\nolimits_2 e^{-\beta(
{\cal H}_0-2\Delta)}-
\frac{1}{2} \left(\mathop{\rm Tr}\nolimits_1 e^{-\beta({\cal H}_0
-\Delta)}\right)^2\right]
\label{b20}
\end{equation}
as the effect of the Bose statistics of the vacancies. The combination
\begin{equation}
b_2=b_2^0+b_2^{\rm int}
\label{b2tot}
\end{equation}
yields the total effect.

The statistical effects are trivial to calculate, as the unperturbed
grand partition function is given by
\FL
\begin{equation}
\ln \Xi^0 =-\sum_{\vec{k}} \ln\left(1-e^{-\beta(\varepsilon(\vec{k})
+\Delta
)}\right) = N \sum_{\ell=1}^{\infty} b_{\ell}^0 e^{-\ell
\beta \Delta},
\label{xi0}
\end{equation}
such that
\begin{equation}
b_{\ell}^0=\frac{1}{N\ell} \sum_{\vec{k}} e^{-\ell\beta
\varepsilon(\vec{k})}
=\frac{1}{\ell} b_1(\beta\ell),
\label{bl0}
\end{equation}
with $b_1(\beta)$ given by (\ref{b1}). Thus it suffices to focus our
attention on the calculation of $b_2^{\rm int}$. As a general
observation
we note that the two contributions in (\ref{b2tot}) will have
opposite
signs. To see this more clearly, we go over to a series in the
density $n$
of the vacancies,
\begin{equation}
n=\frac{1}{N v_0} \langle \sum_i b_i^{\dagger}b_i \rangle =
-\partial p/
\partial \Delta =\frac{1}{v_0} \sum_{\ell=1}^{\infty} \ell b_{\ell}
e^{-\ell \beta
\Delta}.
\label{dens}
\end{equation}
Eliminating $e^{-\beta\Delta}$ from (\ref{vircs}) and (\ref{dens}) we
obtain a virial expansion for the pressure
\begin{equation}
\beta p = n -(b_2/b_1^2) v_0 n^2 + \ldots .
\label{pexp}
\end{equation}
So at fixed density $n$ the statistical effects lower the pressure as
$b_2^0 $ is positive according to (\ref{bl0}). The hard-core repulsion
can only increase the pressure, so $b_2^{\rm int}$ must be negative.

\section{The second virial coefficient}

The second virial coefficient (\ref{b2int}) is evaluated as
\begin{equation}
b_2^{\rm int} =\frac{1}{N} \int dE \left[ \rho_2(E) -\rho_2^0(E)
\right]
e^{-\beta(E-2\Delta)},
\label{b2e}
\end{equation}
where the level densities $\rho_2(E)$ are given by
\begin{equation}
\left\{
\begin{array}{l}
\rho_2(E)=\mathop{\rm Tr}\nolimits_2 \delta(E-{\cal H}) \\
\rho_2^0(E)= \mathop{\rm Tr}\nolimits_2 \delta(E-{\cal H}_0).
\end{array}
\right.
\label{rhos}
\end{equation}
The level densities are obtained from the formula
\begin{equation}
{\cal G}_+(E) = \frac{1}{E+i\epsilon-{\cal H}}={\cal P}
\frac{1}{E-{\cal H}}
-i\pi\delta(E-{\cal H}),
\label{Green}
\end{equation}
which transfers the problem to the determination of the Green's
function,
for which the general equation holds
\begin{equation}
{\cal G}(z) = {\cal G}_0(z)+{\cal G}_0(z) V {\cal G}(z),
\label{Geq}
\end{equation}
with $z$ a complex number. The density $\rho(E)$ is obtained from
${\cal G}$ as
\begin{equation}
\rho(E)=-\frac{1}{\pi} \mathop{\rm Im}\nolimits
\mathop{\rm Tr}\nolimits {\cal G}_+ (E).
\label{rhoG}
\end{equation}

The states of the unperturbed 2-vacancy Hamiltonian are denoted by
two
wavenumbers $\vec{k}_1$ and $\vec{k}_2$ with the property
\begin{eqnarray}
{\cal H}_0 |\vec{k}_1\,\vec{k}_2\rangle &=&E_0(\vec{k}_1, \vec{k}_2)
|\vec{k}_1\,\vec{k}_2\rangle\nonumber\\
&=&\left(2\Delta+\varepsilon(\vec{k}_1)
+\varepsilon(\vec{k}_2) \right) |\vec{k}_1\,\vec{k}_2\rangle.
\label{states}
\end{eqnarray}
The matrix elements of $V$ are obtained from (\ref{vk})
\begin{equation}
\langle\vec{k}_1\,\vec{k}_2| V |\vec{k}_1'\,\vec{k}_2'\rangle =
\frac{2U}{N} \sum_{\vec{G}}
\delta_{\vec{k}_1+\vec{k}_2,\vec{k}_1'+\vec{k}_2'+\vec{G}},
\label{vmat}
\end{equation}
which shows that a representation in center of mass and relative
coordinates will be advantageous. Thus we introduce
\begin{equation}
\vec{K}=\vec{k}_1+\vec{k}_2,\mbox{~~~~~}\vec{k}=(\vec{k}_1-
\vec{k}_2)/2.
\label{kk}
\end{equation}
The sum over $\vec{G}$
is eliminated by choosing the Brillouin zone in such a way that
no two points
in it have values of $\vec{K}$ differing by a reciprocal lattice
vector,
so that only the term with $\vec{G}=\vec{0}$ contributes.
The matrix element of $V$ is then diagonal in $\vec{K}$. From
now on
we will assume that the Brillouin zone has been chosen in such a way,
and drop the reference to $\vec{G}$.
Since the total momentum $\vec{K}$ is conserved by ${\cal H}_0$
and $V$, ${\cal G}$ becomes diagonal in it. So, writing
(\ref{Geq}) in the $\vec{K},\vec{k}$ representation,
we have
\begin{eqnarray}
\langle\vec{K}\,\vec{k}| {\cal G}(z) |\vec{K}\,\vec{k}'\rangle & = &
\frac{1}{z-E_0(\vec{K},\vec{k})} \left\{ \vphantom{|\sum_{3}|}
\delta_{\vec{k},\vec{k}'}\right.
\nonumber \\
&&+\left.
\frac{2U}{N}\sum_{\vec{k}''} \langle\vec{K}\,\vec{k}''| {\cal G}(z)
|\vec{K}\,\vec{k}'\rangle \right\}.\nonumber\\
\label{Gdiag}
\end{eqnarray}
The simplifying feature of (\ref{Gdiag}) is that the general
matrix element
of ${\cal G}$ couples only to the total sum over the first entry
of the
matrix elements. For the latter we obtain an expression by summing
(\ref{Gdiag}) over $\vec{k}$
\begin{eqnarray}
\sum_{\vec{k}}\langle\vec{K}\,\vec{k}| {\cal G}(z&)& |
\vec{K}\,\vec{k}'\rangle =
\frac{1}{z-E_0(\vec{K},\vec{k}')}\nonumber \\
& & +2U{\cal R}(z,\vec{K})  \sum_{\vec{k}''}
\langle\vec{K}\,\vec{k}''| {\cal G}(z)
|\vec{K}\,\vec{k}'\rangle,\nonumber\\
\label{sumk}
\end{eqnarray}
with ${\cal R}(z,\vec{K}) $ given by
\begin{equation}
{\cal R}(z,\vec{K}) =\frac{1}{N} \sum_{\vec{k}}
\frac{1}{z-E_0(\vec{K},
\vec{k})}.
\label{rdef}
\end{equation}
Now (\ref{sumk}) is an algebraic equation for the quantity
in the left hand
side. Using the solution of this equation in (\ref{Gdiag})
one finds
\begin{eqnarray}
\langle\vec{K}\,\vec{k}| {\cal G}(z&)& |\vec{K}\,\vec{k}'\rangle  =
\frac{1}{z-E_0(\vec{K},\vec{k})} \left\{ \vphantom{\frac{1}{2^2}
\sum_{3^4}}
\delta_{\vec{k},\vec{k}'}\right.
\nonumber \\
&& +\left.
\frac{2U/N}{1-2U{\cal R}(z,\vec{K})} \frac{1}{z-E_0(\vec{K},\vec{k}')}
\right\}.\nonumber\\
\label{gsol}
\end{eqnarray}
Note that the off-diagonal elements are of order $N^{-1}$ while the
diagonal elements are of order 1. With (\ref{gsol}) we can
calculate the
level density from
\begin{equation}
\mathop{\rm Tr}\nolimits {\cal G}(z)=\sum_{\vec{K},\vec{k}}
\langle\vec{K}\,\vec{k}| {\cal G}(z)
|\vec{K}\,\vec{k} \rangle,
\label{trace}
\end{equation}
leading to the compact expression
\begin{equation}
\mathop{\rm Tr}\nolimits {\cal G}(z) =\mathop{\rm Tr}\nolimits
{\cal G}_0(z) +\frac{\partial}{\partial z} \sum_{\vec{K}}
\ln \left[ 1-2U{\cal R}(z,\vec{K}) \right] ,
\label{trdz}
\end{equation}
which in turn yields for the difference $\rho_2-\rho_2^0$
\begin{eqnarray}
\rho_2(E)&-&\rho_2^0(E)=\nonumber\\
&-&\frac{1}{\pi}\mathop{\rm Im}\nolimits\left[ \frac{\partial}
{\partial z}
\sum_{\vec{K}}\ln \left( 1
-2U{\cal R}(z,\vec{K})\right) \right]_{z=E+i\epsilon}.\nonumber\\
\label{rhodiff}
\end{eqnarray}

This expression holds generally for any on-site repulsion $U$.
We may let
$U\rightarrow\infty$, by which it will disappear from  the formula
as the $U$ term under the logarithm starts to dominate the argument
for any $z$ and $\vec{K}$. Omitting the 1 in the argument of the
logarithm,
the term $\ln( -2U)$ drops out after differentiation with respect
to $z$.
So for (\ref{rhodiff}) we have in the limit $U\rightarrow\infty$ the
equivalent expression
\begin{equation}
\rho_2(E)-\rho_2^0(E)=-\frac{1}{\pi}\mathop{\rm Im}\nolimits\left[
\frac{\partial}{\partial z}
\sum_{\vec{K}}\ln {\cal R}(z,\vec{K}) \right]_{z=E+i\epsilon}.
\label{rhodiff2}
\end{equation}
Hereby the problem is essentially reduced to the evaluation of
${\cal R}
(z,\vec{K})$ given by (\ref{rdef}).

A few comments are in order about this expression. For bosons
the state
$|\vec{k}_1\,\vec{k}_2\rangle $ is the same as
$|\vec{k}_2\,\vec{k}_1\rangle$.
So the relative momenta $\vec{k}$ and $-\vec{k}$ should be
identified
with each other. Both $\vec{k}_1$ and $\vec{k}_2$ run through
a Brillouin zone
appropriate for the structure of the lattice, and this in
principle
defines the ranges of $\vec{K}$ and $\vec{k}$. But as was
mentioned before,
we have chosen the Brillouin zone such that there are no
points whose values
of $\vec{K}$ differ by a reciprocal lattice vector.

{}From the definition (\ref{rdef}) of ${\cal R}(z,\vec{K})$ and the
expression (\ref{states}) for $E_0(\vec{K},\vec{k})$ one sees
that $2\Delta$
occurs as a shift in the energy variable and the $\rho_2$ are
functions of
$E-2\Delta$. Taking $E-2\Delta$ as an integration variable in
(\ref{b2e})
one finds
\begin{equation}
b_2^{\rm int} =\frac{1}{N} \int dE' \left[ \rho_2(E') -
\rho_2^0(E')\right]
e^{-\beta E'},
\label{b2e2}
\end{equation}
with
\FL
\begin{equation}
\rho_2(E')-\rho_2^0(E')= -\frac{1}{\pi}\mathop{\rm Im}\nolimits
\left[ \frac{\partial}{\partial z}
\sum_{\vec{K}} \ln {\cal R}'(z,\vec{K}) \right]_{z=E'+i\epsilon} ,
\label{rhoe2}
\end{equation}
and ${\cal R}'(z,\vec{K})$ given by
\begin{equation}
{\cal R}'(z,\vec{K}) = \frac{1}{N} \sum_{\vec{k}} \frac{1}
{z-\varepsilon(
\frac{\vec{K}}{2}+\vec{k}) -\varepsilon(\frac{\vec{K}}{2} -\vec{k})} .
\label{r2}
\end{equation}
In this formula $\Delta$ is eliminated, as it should be. Using the
fact that $\mathop{\rm Im}\nolimits \ln z = \mathop{\rm arg}\nolimits
z$, we can further rewrite (\ref{rhoe2})
\begin{equation}
\rho_2(E') - \rho_2^0(E') = -\frac{N}{\pi} \frac{\partial}
{\partial z}
F(z) |_{z=E'+i\epsilon},
\label{rhof}
\end{equation}
with
\begin{equation}
F(z)=\frac{1}{N} \sum_{\vec{K}} \mathop{\rm arg}\nolimits
{\cal R}'(z,\vec{K}).
\label{fdef}
\end{equation}

{}From (\ref{rhoe2}) and (\ref{r2}) one sees that only $E'$
values occur
in (\ref{b2e2}) which lead to complex values of
${\cal R}'(z,\vec{K})$.
These occur when $z=E'+i\epsilon$ is a pole in the
$\vec{k}$-integration
in (\ref{r2}). Thus the combined bandwidth of $\varepsilon(
\vec{K}/2+\vec{k}) +\varepsilon(\vec{K}/2 -\vec{k}) $
determines the range of $E'$ values. This is twice the
bandwidth of
$\varepsilon(\vec{k})$ just as in the ideal Bose contribution
(\ref{bl0}).

\section{The one-dimensional case}
\label{oned}

We interrupt the general discussion for the treatment of the
one-dimensional case of (\ref{b2e2})-(\ref{fdef}), as this case is
interesting, completely analyzable, and elucidating for the structure
of the functions occurring in (\ref{b2e2})-(\ref{fdef}).

The $d=1$ band structure is given by (for a lattice constant $a=1$)
\begin{equation}
\varepsilon(k)=2t(1-\cos k),
\label{1deps}
\end{equation}
which gives for $b_1(\beta)$ the expression
\begin{equation}
b_1(\beta)=\frac{1}{2\pi} \int_{-\pi}^{\pi} dk \,
e^{-2\beta t (1-\cos k)}=
e^{-2\beta t} I_0(2 \beta t),
\label{1db1}
\end{equation}
where $I_0(x)$ is a modified Bessel function of the first kind.

For the evaluation of ${\cal R}(z,K)$ we rearrange the
Brillouin zone
in such a way that its boundaries in $K,k$ space are
convenient, and also such that no reciprocal lattice vector
$G$ enters
into the problem. In figure \ref{bzfig} we have divided the
original
Brillouin zone $-\pi < k_1 < \pi\;,\; -\pi < k_2 < \pi$ into
four domains,
I, II, III, and IV. Domains I and IV, as well as II and III, refer
to the same states,
as they are obtained from each other through interchanging $k_1
\leftrightarrow k_2$. So it suffices to take one of each,
say I and III.
Now III is equivalent to III$'$, which follows from III by shifting
$k_1$ over $2\pi$. The combined domain I and III$'$ is given in $K,
k$ space by $0 < K < 2\pi\;,\; 0 < k < \pi$.

Using this parameter space one can write (\ref{r2}) explicitly as
(dropping the primes)
\begin{equation}
{\cal R}(z,K)=\frac{1}{2\pi} \int_0^{\pi} dk\,
\frac{1}{z-4t(1-\cos\frac{K}{2}
\cos k)} .
\label{1dr}
\end{equation}
For convenience we put $z=4t(\zeta+1+i\epsilon)$, and have
\begin{equation}
{\cal R}(\zeta,K)=\frac{1}{8 \pi t}\int_0^{\pi} dk \,\frac{1}
{\zeta
+i\epsilon+\cos
\frac{K}{2} \cos k}.
\label{1dr2}
\end{equation}
According to equations (\ref{rhof}) and (\ref{fdef}) we are
interested
in real values of $\zeta$. So, letting $\epsilon \downarrow 0$,
the integral has a real part given by a principal value integral,
and an imaginary part given by an integral over a delta function
\begin{eqnarray}
{\cal R}(\zeta,K)&=& \frac{1}{8 \pi t} \,{\cal P}\int_0^{\pi} dk\,
\frac{1}
{\zeta+\cos
\frac{K}{2} \cos k}\nonumber\\
&&-\frac{i}{8t}\int_0^{\pi} dk\, \delta(\zeta+\cos
\frac{K}{2} \cos k).
\label{rint}
\end{eqnarray}
Two cases must be distinguished: for $|\zeta| > |\cos(K/2)|$, the
imaginary part of ${\cal R}$ is zero, and for $|\zeta| <
|\cos(K/2)|$
its real part is zero. The result is
\FL
\begin{equation}
{\cal R}(\zeta,K)=\left\{
\begin{array}{ll}
\displaystyle{\frac{1}{8t} \frac{\mathop{\rm sgn} (\zeta)}
{\sqrt{\zeta^2-\cos^2\frac{K}{2}}}}&
{}~~(|\zeta| > |\cos(K/2)|)\\
\\
\displaystyle{\frac{-i}{8t} \frac{1}{\sqrt{\cos^2\frac{K}{2}
-\zeta^2}}}&
{}~~(|\zeta| < |\cos(K/2)|).
\end{array}
\right.
\label{rintsecond}
\end{equation}
The important quantity in
equations (\ref{rhof}) and (\ref{fdef}) is $\arg {\cal R}$. From
(\ref{rintsecond}) we see that
if $\zeta < -|\cos(K/2)|$, ${\cal R}$
is real and negative, while for $\zeta > |\cos(K/2)|$ it is real and
positive. For $-|\cos(K/2)| \leq \zeta \leq |\cos(K/2)|$,
${\cal R}$ is pure imaginary with a negative imaginary part.
So we find that
\FL
\begin{equation}
\Phi(\zeta,K)=\mathop{\rm arg}\nolimits {\cal R}(\zeta,K)=\left\{
\begin{array}{ll}
-\pi & ~~(\zeta < -|\cos(K/2)|) \nonumber\\
-\pi/2 & ~~(|\zeta| \leq |\cos(K/2)|)\\
0 & ~~(\zeta > |\cos(K/2)|).\nonumber
\end{array}
\right.
\label{phi}
\end{equation}
These values of $\Phi$ are plotted in figure \ref{fifig} in the
$\zeta,
K/2$ plane.
Integrating in the $K$-direction, we find for $F(\zeta)$
\begin{eqnarray}
F(\zeta)&=&\frac{1}{N} \sum_K \mathop{\rm arg}\nolimits
{\cal R}(\zeta, K) = \frac{1}{2\pi}
\int_0^{2\pi} dK \Phi(\zeta,K) \nonumber \\
&&~~~= \left\{
\begin{array}{ll}
-\pi & ~~(\zeta < -1) \\
- \arccos \zeta & ~~(-1 \leq \zeta \leq 1)\\
0 & ~~( \zeta > 1).
\end{array}
\right.
\label{1df}
\end{eqnarray}
Using $\zeta$ instead of $E$ in (\ref{b2e2}) as integration variable
we have
\begin{equation}
b_2^{\rm int} = \frac{1}{N}  \int_{-1}^1 d\zeta \left[
\rho_2(\zeta) - \rho_2^0(\zeta)\right] e^{-4\beta t(\zeta +1)} .
\label{rhozeta}
\end{equation}
Due to the differentiation with respect to $z$ in (\ref{rhof})
only values of $\zeta$ between $-1$ and 1 contribute to the
integral.
{}From (\ref{1df}) we see that for $|\zeta| \leq 1$
\begin{equation}
\rho_2(\zeta) - \rho_2^0(\zeta)= -\frac{N}{\pi} \frac{\partial}
{\partial
 \zeta} F(\zeta)=
-\frac{N}{\pi} \frac{1}{\sqrt{1-\zeta^2}}.
\label{rho1d}
\end{equation}
Substituting this into (\ref{rhozeta}) we find
\begin{equation}
b_2^{\rm int}=-e^{-4\beta t} I_0(4\beta t).
\label{1dbint}
\end{equation}

We now compare this with the result for an ideal lattice gas,
given by
(\ref{bl0})
\begin{equation}
b_2^0=\frac{1}{2}b_1(2\beta)=\frac{1}{2}e^{-4 \beta t}
I_0(4\beta t)=
-\frac{1}{2} b_2^{\rm int}.
\label{1db20}
\end{equation}
So the hard-core interaction giving rise to $b_2^{\rm int}$
changes the
value of $b_2$ from $b_2^0$ into its opposite, or in other words,
the Bose value is turned into the Fermi value. This is exactly
what
has to be expected from the well known fact that a hard-core
Bose gas
in one dimension is equivalent to an ideal Fermi gas (for all
virial
coefficients).

The picture shown in figure \ref{fifig} for the phase $\Phi$
has some
general validity, in the sense that for sufficiently negative
$\zeta$ one has $\Phi =-\pi$ while for $\zeta$ sufficiently
positive
$\Phi=0$. In the zone in between one has for $d>1$ in general a
continuous transition from $-\pi$ to 0, with eventually also zones
with $\Phi=-\pi/2$.

\section{The two- and three-dimensional lattices}

The calculation of $b_2$ for hypercubic lattices
in higher dimensions runs along the same lines
as what was done in section \ref{oned} for one dimension.
The band structure is now given by (again $a=1$)
\begin{equation}
\varepsilon (\vec{k}) =\left\{
\begin{array}{l}
2t(2-\cos k_x - \cos k_y ) ~~~~~~~~~~~~~~~~(d=2)\\
2t(3-\cos k_x - \cos k_y -\cos k_z) ~~~~~(d=3).
\end{array}
\right.
\label{hdeps}
\end{equation}
In $d$ dimensions, $b_1(\beta)$ is simply given by
\begin{eqnarray}
b_1(\beta) &=& \frac{1}{(2\pi)^d} \left[ \int_{-\pi}^{\pi}
dk\, e^{-2\beta t(1-\cos k)}\right]^d\nonumber\\
&=&e^{-2d\beta t}\left[ I_0
(2\beta t) \right]^d.
\label{mdb1}
\end{eqnarray}

As before, we rearrange the Brillouin zone such that its
boundaries in
$\vec{K},\vec{k}$ space are convenient. The different
components of
$\vec{K},\vec{k}$ are independent, and we have $0\leq K_i
\leq 2\pi\;,
\; 0 \leq k_x \leq \pi \; , \; -\pi \leq k_y ,
k_z \leq \pi$. The $k_x$-interval is halved to avoid
counting the same (symmetric) state twice.
The two-vacancy energy bands are given by
\begin{eqnarray}
E_0(\vec{K},\vec{k})&=&
4t(2-\cos\displaystyle{\frac{K_x}{2}}\cos k_x -\cos
\displaystyle{\frac{K_y}{2}}\cos k_y )\nonumber\\
&&~~~~~~~~~~~~~~~~~~~~~~~~~~~~~~~~~~~~~~~(d=2),\nonumber\\
E_0(\vec{K},\vec{k})&=&
4t(3-\cos\displaystyle{\frac{K_x}{2}}\cos k_x -\cos
\displaystyle{\frac{K_y}{2}}\cos k_y
\nonumber \\
&&~~~~~~~~~~~-\cos\frac{K_z}{2}\cos k_z)~~~~~~(d=3).
\nonumber\\
\label{md2es}
\end{eqnarray}
We scale the parameter $z$ as
\begin{equation}
z=\left\{
\begin{array}{l}
\hphantom{1}8t(\zeta +1 +i\epsilon)~~~~~~(d=2)\\
12t(\zeta +1 +i\epsilon)~~~~~~(d=3),
\end{array}
\right.
\end{equation}
so that in both cases the energy band runs from $\zeta=-1$
to $\zeta=1$.

In two dimensions we find with (\ref{rint})
\widetext
\begin{eqnarray}
{\cal R}_2(\zeta,\vec{K})&=&\frac{1}{(2\pi)^2}
\int_{-\pi}^{\pi}dk_y\int_
{0}^{\pi} dk_x \, \frac{1}{8t} \left[\zeta+i\epsilon
+ \frac{1}{2}\left(
\cos\frac{K_x}{2}\cos k_x
+\cos\frac{K_y}{2}\cos k_y \right)
\right]^{-1}
\nonumber\\
&=& \frac{1}{32t\pi} \int_{-\pi}^{\pi}dk_y \frac{f(A,B)}{
\sqrt{|A^2-B^2|}},
\label{2dr}
\end{eqnarray}
\narrowtext
where
\begin{eqnarray}
A&=& \zeta+\frac{1}{2} \cos\frac{K_y}{2}\cos k_y,\nonumber\\
B&=& \frac{1}{2}\cos\frac{K_x}{2},
\label{AB}
\end{eqnarray}
and
\begin{equation}
f(A,B)=\left\{
\begin{array}{ll}
\mathop{\rm sgn} (A) &~~(A^2>B^2)\\
-i &~~(A^2<B^2).
\end{array}
\right.
\label{f}
\end{equation}
The integral in (\ref{2dr}) can be expressed in terms of complete
elliptic integrals of the first kind (see appendix). This gives
an analytic expression for ${\cal R}_2(\zeta,\vec{K})$.

This expression can also be used to find the result in three
dimensions,
\widetext
\begin{equation}
{\cal R}_3(z,\vec{K})=\frac{1}{(2\pi)^3} \int_{-\pi}^{\pi} dk_z
\int_{-\pi}^{\pi} dk_y \int_0^{\pi} dk_x\left[
z-4t(3-\cos\frac{K_x}{2}\cos k_x
-\cos\frac{K_y}{2}\cos k_y
-\cos\frac{K_z}{2}\cos k_z)\right]^{-1}
\label{3drz}
\end{equation}
or
\begin{eqnarray}
{\cal R}_3(\zeta,\vec{K})&=&\frac{1}{2\pi} \int_{-\pi}^{\pi} dk_z
\frac{1}{32t\pi} \int_{-\pi}^{\pi}dk_y\int_0^{\pi} dk_x \left[
\widetilde{\zeta} +i\epsilon \vphantom{\frac{1}{2}}
+\frac{1}{2}(
\cos\frac{K_x}{2}\cos k_x +\cos\frac{K_y}{2}\cos k_y)\right]^{-1}
 \nonumber\\
&=& \frac{1}{2\pi}\int_{-\pi}^{\pi} dk_z {\cal R}_2(
\widetilde{\zeta},K_x,
K_y),
\label{r3}
\end{eqnarray}
\narrowtext
where
\begin{equation}
8t(\widetilde{\zeta} +1)=12t(\zeta+1)-4t(1-\cos\frac{K_z}{2}
\cos k_z).
\label{tilde}
\end{equation}
Using the analytic expression found for ${\cal R}_2$,
${\cal R}_3$ can be
calculated by numerical integration of (\ref{r3}).

The next step is to obtain $F(\zeta)$ as given in (\ref{fdef}) by
integrating over $\vec{K}$. This is done numerically
using Monte Carlo integration. The result is a function $F(\zeta)$
that is equal to $-\pi$ for $\zeta < -1$, where ${\cal R}_d$ is
real and negative, equal to $0$ for $\zeta > 1$, where ${\cal R}_d$
is real and positive, and that is in between these two values for
$\zeta \in [-1,1]$, where ${\cal R}_d$ is complex. It follows from
the symmetry of ${\cal R}_d$ that
\begin{equation}
F(-\zeta)=-\pi-F(\zeta).
\label{sym}
\end{equation}
Figures \ref{f2} and \ref{f3} show $F(\zeta)$ for the square and
cubic lattices.

Using $F(\zeta)$, $b_2^{\rm int}$ can be found from equations (\ref
{b2e2}) and (\ref{rhof})
\begin{equation}
b_2^{\rm int}=-\frac{1}{\pi}\int_{-1}^{1} d\zeta \left(
\frac{\partial}{\partial \zeta} F(\zeta)\right) e^{-4d\beta
t(\zeta +1)}.
\label{mdb2}
\end{equation}
Partially integrating, we find
\begin{eqnarray}
b_2^{\rm int}&=&-e^{-4d\beta t} \left\{
e^{4d\beta t}+\frac{4d\beta t}{\pi} \int_{-1}^{1} d\zeta
F(\zeta) e^{-4d\beta t\zeta}\right\}\nonumber\\
&=&-e^{-4d\beta t} \left\{1-
\frac{8d\beta t}{\pi} \int_0^{1} d\zeta
F(\zeta) \sinh{4d\beta t\zeta}\right\}.\nonumber\\
\end{eqnarray}
To find $b_2^{\rm int}$, the last equation can be numerically
integrated
for various values of $\beta t$.

As before, we compare this with the result for the ideal lattice
gas (\ref{bl0})
\begin{equation}
b_2^0=\frac{1}{2}b_1(2\beta)=\frac{1}{2}e^{-4d\beta t}
[I_0(4\beta t)]^d.
\label{mdb20}
\end{equation}
The ratio $b_2^{\rm int}/b_2^0$ is given by
\FL
\begin{equation}
b_2^{\rm int}/b_2^0=
\frac{-2}{[I_0(4\beta t)]^d}\left\{1-\frac{8d\beta t}{\pi}
\int_0^1d\zeta
F(\zeta) \sinh 4d\beta t\zeta\right\}.
\label{ratio}
\end{equation}
For $\beta t \rightarrow 0$, this ratio goes to $-2$. The
second virial
coefficient $b_2=b_2^0+b_2^{\rm int}$ thus approaches $-b_2^0$
for high temperatures.
This is indicative of fermionic behaviour, which is indeed what
one would expect: at high temperatures, the only important
contribution
to the free energy of the system is the entropy involved in
distributing
a certain number of hard core particles over the lattice. This
is the
same as for fermions. For $\beta t \rightarrow \infty$
the behaviour of (\ref{ratio}) depends on the behaviour of
$F(\zeta)$
for $\zeta\rightarrow 1$. In two dimensions, $F(\zeta)\propto
(1-\zeta)/\ln(1-\zeta)$, which leads to $b_2^{\rm int}/b_2^0$
going to
zero like $-1/\ln(\beta t)$ as $\beta t \rightarrow \infty$.
The same low-temperature behaviour is found for quantum hard disks
\cite{Schick}.
In three dimensions, $F(\zeta) \propto (1-\zeta)^2$, which gives
$b_2^{\rm int}/b_2^0 \propto -1/\sqrt{\beta t}$ for $\beta t
\rightarrow
\infty$, just as for quantum hard spheres \cite{Uhl}.
So in both cases, $b_2$
approaches the value for the ideal Bose gas at low temperatures.
This
shows that when the thermal wavelength (\ref{lambda}) exceeds
the lattice constant $a(=1)$, the effects of the Bose statistics
starts to dominate.
Plots of $b_2/b_2^0$ versus $\beta t$ for two and three
dimensions are given in figures \ref{rat2} and \ref{rat3}.

\section{Conclusion}
The pressure and density of a gas of quantum particles on a lattice
can be expressed in fugacity expansions,
\begin{eqnarray}
p&=&\frac{1}{\beta v_0} \sum_{\ell=1}^{\infty} b_{\ell}
e^{-\ell\beta\Delta},
\label{pee}\\
n&=&\frac{1}{v_0} \sum_{\ell=1}^{\infty} \ell b_{\ell}
e^{-\ell\beta\Delta}.
\label{en}
\end{eqnarray}
For bosons with a hard-core interaction, the first two coefficients,
$b_1$ and $b_2$, can be calculated. It turns out that the
effect of the
hard-core interaction on $b_2$ depends strongly on temperature and
on the transfer integral $t$. For small $\beta t$ it is fermionic in
character, as far as $b_2$ is concerned. For large $\beta t$
the effect
of the hard-core disappears, and only the Bose character remains.

The system of vacancies that exists in solid $^4$He should
behave to a good approximation like this simple model: the
vacancies
move through the crystal lattice by a tunneling process,
they are
bosons, and they have a hard-core repulsion. The gas of
vacancies
has been experimentally studied by probing the attenuation
and the
velocity shift of sound in a $^4$He crystal \cite{Goodk}.
Very pure
hcp Helium was used, so that the effects of both the phonons
and the
delocalized, bosonic vacancies could be observed. Treating the
vacancies
as a gas of free particles, it was seen that they obey Bose
statistics.
For the expressions (\ref{pee}) and (\ref{en}) this means that
not only
the first term, but at least also the second is experimentally
observable.
One can use the expressions (\ref{b1l}) and (\ref{bl0})
for a gas of
free, ideal bosons to estimate the order of magnitude of
the various
terms in (\ref{pee}) and (\ref{en}). One finds that $b_{\ell}^0=
v_0/\ell^{5/2}\lambda^3$, so that
\begin{eqnarray}
p&=&\frac{1}{\beta \lambda^3} \sum_{\ell=1}^{\infty}
\frac{e^{-\ell\beta\Delta}}{\ell^{5/2}},
\label{peef}\\
n&=&\frac{1}{\lambda^3} \sum_{\ell=1}^{\infty}
\frac{e^{-\ell\beta\Delta}}{\ell^{3/2}}.
\label{enf}
\end{eqnarray}
Using the value $\Delta/k_B=0.71 {\rm K}$ found in \cite{Goodk},
and the
maximum temperature $T= 0.85 {\rm K}$ at which the experiments
were
performed, this shows that the ratio between the second and
first terms
in (\ref{enf}) is $0.15$, and that between the third and first
terms
is $0.04$.

If experiments can be done that detect the contribution of $15\%$
of the second term in (\ref{enf}), it is also possible to detect
the
effects of the hard-core on the coefficient $b_2$ in (\ref{en}).
As can be seen from figure~\ref{rat3}, it varies considerably with
$\beta t$, and it can even change sign. However, it is difficult
to extract this information from the attenuation experiment, since
there
it is not clear what exactly the relation between the measured
quantities and the vacancy density is. It would be necessary to
directly measure, say, the pressure due to the vacancies
\cite{Steve}, and
then compare this with (\ref{pee}). In such an experiment
it would be
crucial to take the hard-core effects into account.
Work is in progress to calculate
the coefficient $b_2$ for the hcp lattice; however, the results are
not expected to differ much from those given for the simpler cubic
lattice here.

The results for the two-dimensional system might have relevance
for low-density $^4$He films on textured substrates \cite{Rehr}.
There, the presence of adsorption sites localizes the Helium
particles on the points of a two-dimensional lattice, defined by
the substrate. Thus the tight-binding approach used in this paper
becomes applicable. The adsorption sites cannot hold more than
one particle, so that the hard-core repulsion is also present.

\acknowledgements

The authors thank J.~Goodkind for making available some of
his experimental data, and for useful discussions;
useful discussions with A.F.~Andreev and S.~Steel
are also acknowledged. Part of this research was supported by
the Natural Sciences and Engineering Research Council of Canada, and
the `Stichting voor Fundamenteel Onderzoek der Materie' (FOM),
which is financially supported by the `Stichting Nederlands
Wetenschappelijk Onderzoek' (NWO).

\unletteredappendix{}

The integral (\ref{2dr}) can be expressed in terms of elliptic
integrals as follows \cite{Mor,Mont}. ${\cal R}_2$ is equal to
\begin{eqnarray}
{\cal R}_2(\zeta,\vec{K})&=&\frac{1}{16t\pi}\int_{A^2>B^2} dk_y
\, \frac{\mathop{\rm sgn} (A)}{\sqrt{A^2-B^2}}\nonumber\\
&&-\frac{i}{16t\pi}\int
_{A^2<B^2} dk_y \frac{1}{\sqrt{B^2-A^2}},
\label{twoint}
\end{eqnarray}
where $k_y\in [0,\pi]$.
On making the substitution $p=\cos k_y$, and writing
\begin{equation}
A^2-B^2=\frac{\cos^2 (K_y/2)}{4}(q_1-p)(q_2-p),
\label{ab}
\end{equation}
where
\begin{equation}
q_1=\frac{-2\zeta + \cos (K_x/2)}{\cos (K_y/2)}, ~~~~~
q_2=\frac{-2\zeta - \cos (K_x/2)}{\cos (K_y/2)},
\label{qs}
\end{equation}
we end up with integrals of the type
\begin{equation}
\int dp \frac{1}{\sqrt{|w|}},
\label{wint}
\end{equation}
where $w=(1-p)(1+p)(q_1-p)(q_2-p)$ and $p\in [-1,1]$. By defining
\begin{equation}
q_-=\min (q_1,q_2),~~~~~q_+=\max (q_1,q_2),
\label{+-}
\end{equation}
we see that values of $p \in [q_-,q_+]$ give $A^2 < B^2$,
and thus contribute to the imaginary part of ${\cal R}_2$.
Values of $p$ outside this interval give $A^2 > B^2$ and thus
contribute to the real part,
for $p<q_-$ with $\mathop{\rm sgn} (A)=-\mathop{\rm sgn}
(\cos(K_y/2))$, and
for $p>q_+$ with $\mathop{\rm sgn} (A)=\mathop{\rm sgn}
(\cos(K_y/2))$.
The result is a sum of integrals of the type (\ref{wint}) between
limits that are consecutive zeroes of $w$. These integrals
can all be expressed in terms of the complete elliptic
integral of the first kind, $K(x)$ \cite{G+R},
\begin{equation}
K(x)=\int_0^1 dp \frac{1}{\sqrt{(1-p^2)(1-x^2p^2)}}
{}~~~~~~(x\in [0,1]).
\label{ka}
\end{equation}
There are several cases to be considered:
\subsection{$q_-,q_+ < -1$ (1a) or $q_-,q_+ > 1$ (1b)}
\begin{equation}
{\cal R}_2(\zeta,\vec{K}) =\frac{\epsilon}{4t\pi\cos(K_y/2)}
\frac{K(r_1)}{\sqrt{(q_+-1)(q_-+1)}} ,
\label{one}
\end{equation}
where $\epsilon=+$ for (1a) and $\epsilon =-$ for (1b).
\subsection{$q_-<-1, q_+>1$ or $q_-,q_+ \in [-1,1]$}
\begin{equation}
{\cal R}_2(\zeta,\vec{K}) =\frac{-i}{4t\pi|\cos(K_y/2)|}
\frac{K(r_2)}{\sqrt{(q_++1)(1-q_-)}} .\label{two}
\end{equation}
\subsection{$q_-<-1,q_+\in[-1,1]$ (3a)\\ or $q_-\in[-1,1],
q_+ > 1$ (3b)}
\begin{eqnarray}
{\cal R}_2(\zeta,\vec{K}) &=&\frac{\epsilon}{4t\pi\cos(K_y/2)}
\frac{K(1/r_1)}{\sqrt{2(q_+-q_-)}}\nonumber\\
&&
-\frac{i}{4t\pi|\cos(K_y/2)|}
\frac{K(1/r_2)}{\sqrt{2(q_+-q_-)}},
\label{three}
\end{eqnarray}
where $\epsilon=+$ for (3a) and $\epsilon =-$ for (3b).
In the above, $r_1$ and $r_2$ are given by
\FL
\begin{equation}
r_1=\sqrt{\frac{2(q_+-q_-)}{(q_+-1)(q_-+1)}},~~~~~
r_2=\sqrt{\frac{2(q_+-q_-)}{(q_++1)(1-q_-)}}.
\label{rs}
\end{equation}

\figure{The Brillouin zone for the one-dimensional problem.
Domains
I and III' are the Brillouin zone used in the calculation.
\label{bzfig}}
\figure{The phase $\Phi(\zeta, K)$ in the $\zeta,K/2$ plane.
\label{fifig}}
\figure{The function $F(\zeta)$ for the square lattice.
The only values of interest are those between $\zeta=-1$ and
$\zeta=1$, where $F(\zeta)$ is not constant and varies from $-\pi$
to 0. Only positive values of $\zeta$ are shown, since $F(\zeta)$
for negative $\zeta$ can be found using the antisymmetry of that
function around $\zeta=0$ and $F(\zeta)=-\pi/2$. \label{f2}}
\figure{The function $F(\zeta)$ for the cubic lattice. \label{f3}}
\figure{The ratio $b_2/b_2^0$ as a function of $\beta t$ for the
square
lattice. For high temperatures (small $\beta t$), $b_2=-b_2^0$,
which is the same value as for fermions. For lower temperatures its
value crosses over to $b_2=b_2^0$, the ideal boson value.
\label{rat2}}
\figure{The ratio $b_2/b_2^0$ as a function of $\beta t$ for
the cubic
lattice. Note that the cross-over to the ideal boson value
is much faster than for the square lattice.\label{rat3}}


\begin{references}
\bibitem{And1} A.F.~Andreev, in: {\em Progress in Low Temperature
Phy\-sics}, Vol.\ 8 (D.F.~Brewer, ed.), North Holland,
Amsterdam, 1982.
\bibitem{And2} A.F.~Andreev and I.M.~Lifshitz, Zh.\ Exp.\
Teor.\ Fiz.\
{\bf 56} 2057 (1969) [Sov.\ Phys.\ JETP {\bf 29} 1107 (1969)].
\bibitem{M+M} T.~Matsubara and H.~Matsuda,
Prog. Theor. Phys. {\bf 16} 569 (1956); H.~Matsuda and T.~Matsubara,
Prog. Theor. Phys. {\bf 17} 19 (1957).
\bibitem{Goodk} G.A.~Lengua and J.M.~Goodkind, J.\ Low Temp.\ Phys.\
{\bf 79} 251 (1990).
\bibitem{Huang} K.~Huang, {\em Statistical Mechanics} (2nd edition),
Wiley, New York, 1987.
\bibitem{Schick} R.L.~Siddon and M.~Schick, Phys.\ Rev.\ A {\bf 9}
907 (1974); W.G.~Gibson, Mol.\ Phys.\ {\bf 49} 103 (1983).
\bibitem{Uhl} G.E.~Uhlenbeck and E.~Beth, Physica {\bf 3} 729 (1936).
\bibitem{Steve} S.~Steel, to be published
\bibitem{Rehr} J.J.~Rehr and M.~Tejwani, Phys.\ Rev.\ B {\bf 20}
345 (1979).
\bibitem{Mor} T.~Morita and T.~Horiguchi, J.\ Math.\ Phys.\ {\bf 12}
986 (1971).
\bibitem{Mont} E.W.~Montroll, in: {\em Proceedings of the third
Berkeley
Symposium on Mathematical Statistics and Probability}, Vol.\ 3
(J.~Neyman, ed.), University of California Press, Berkeley, 1955.
\bibitem{G+R} I.S.~Gradshteyn and I.M.~Ryzhik, {\em Table of
Integrals, Series,
and Products}, Academic Press, San Diego, 1980.

\end{references}
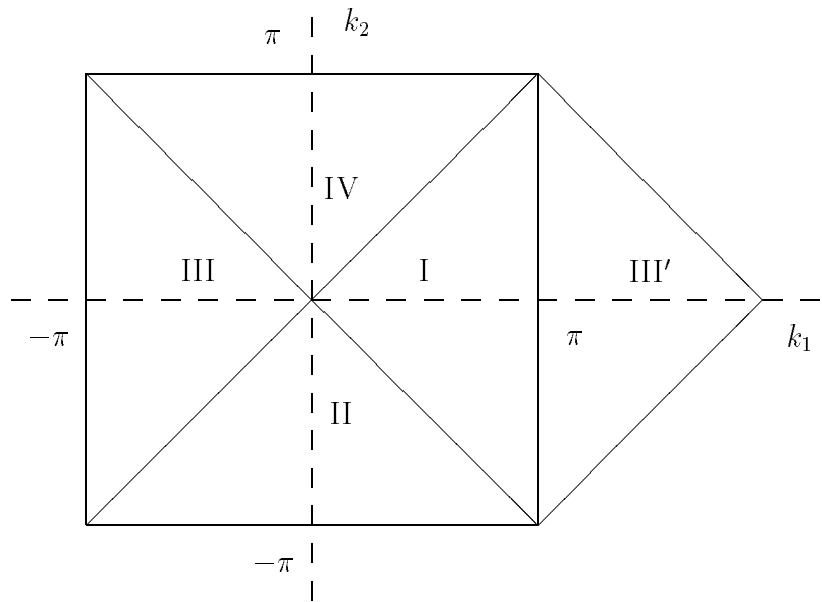
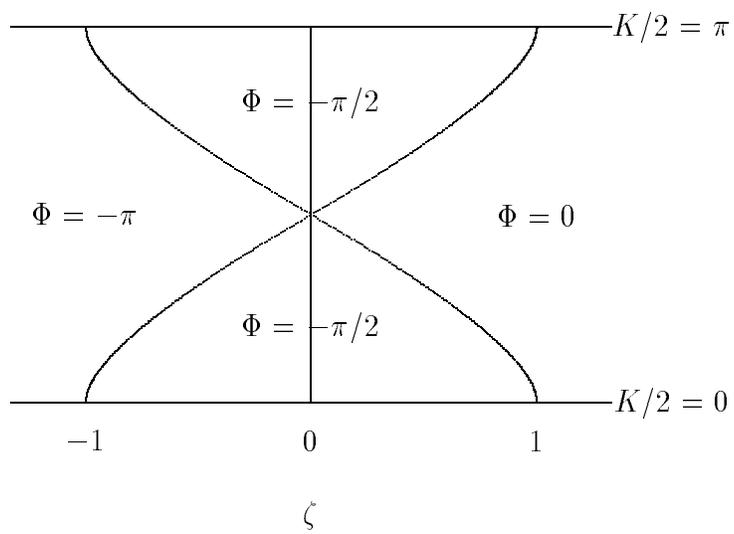
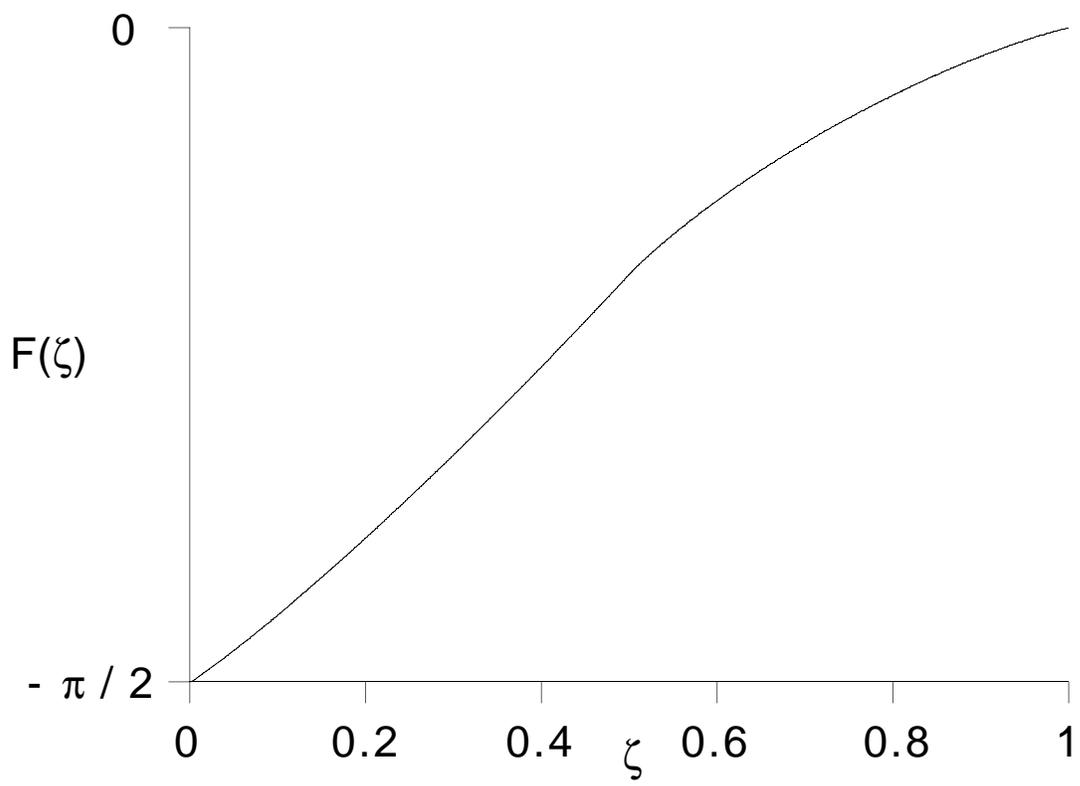
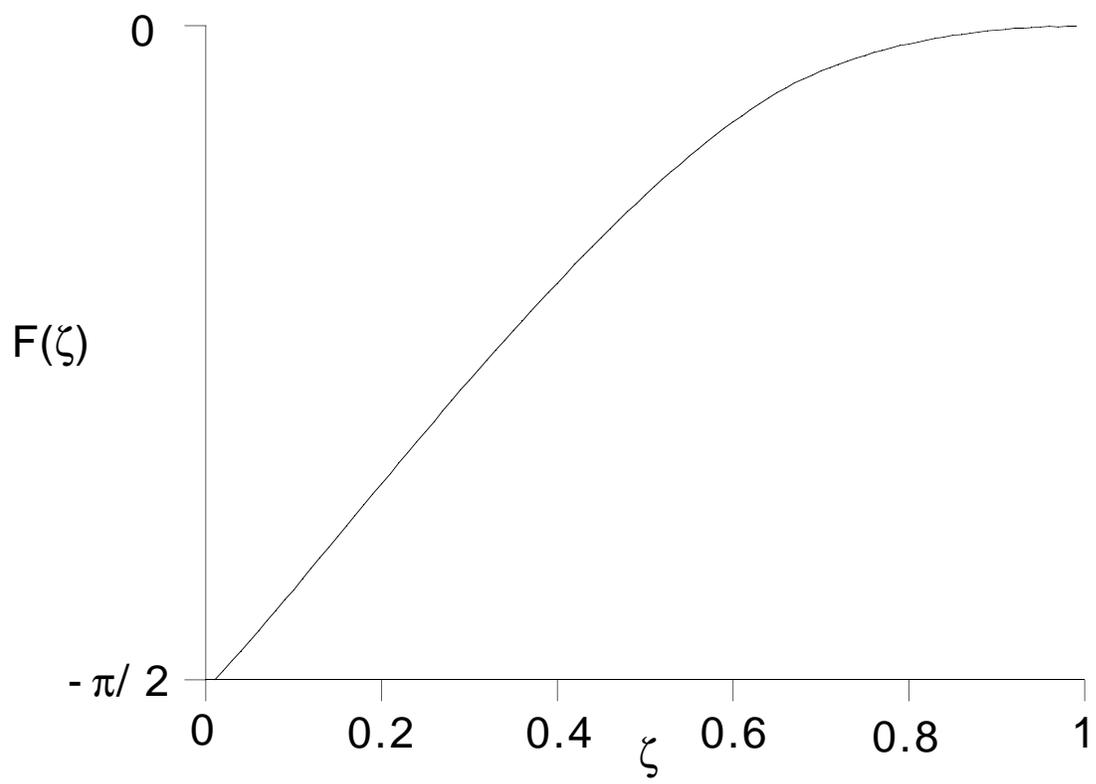
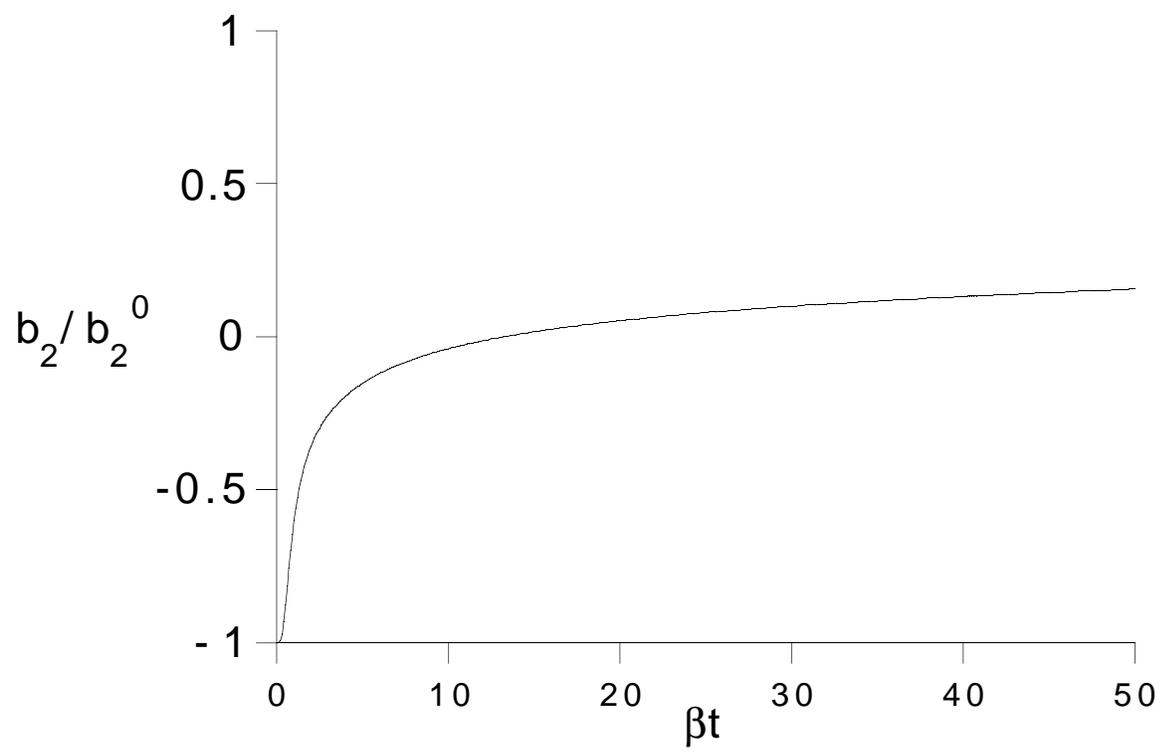
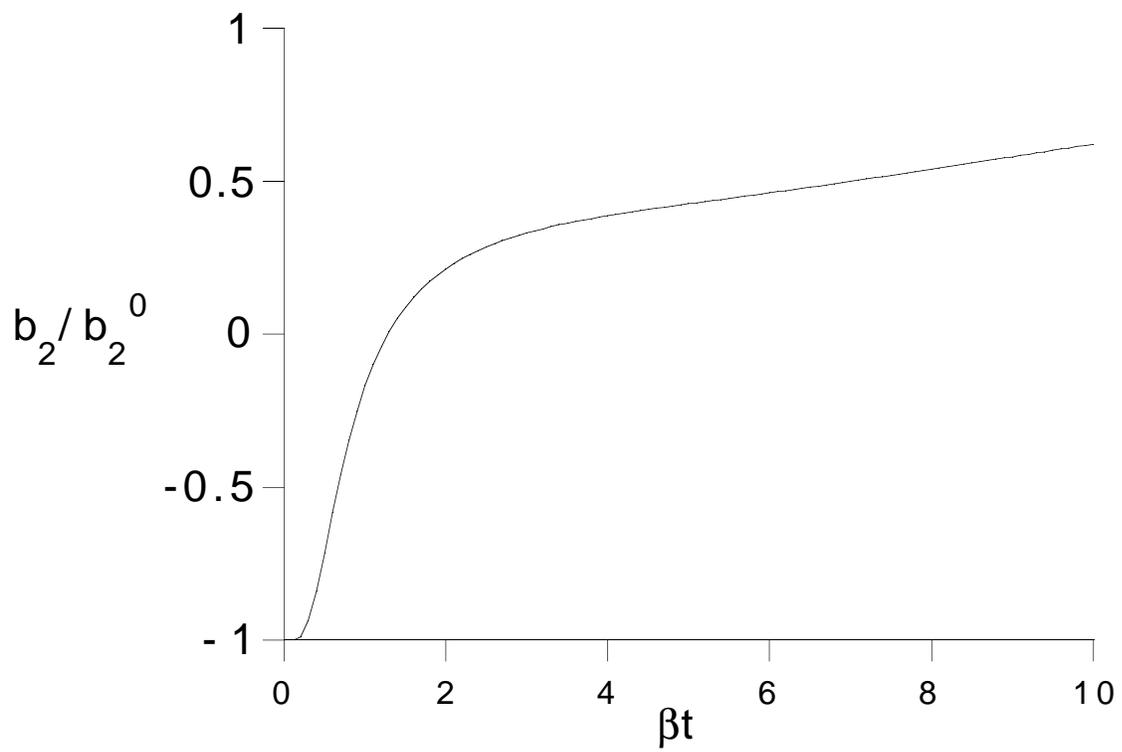
\end{document}